# Dynamic Evolutionary Game Analysis of How Fintech in Banking Mitigates Risks in Agricultural Supply Chain Finance

Qiang Wan[1], Jun Cui[2, *]

1 GuangDong RiZhao Electric Co., Ltd., Huizhou 516100, China; Department of Lifelong Learning, Graduate School, Hanseo University, Seosan-si 31962, Republic of Korea

2 Solbridge International School of Business, Woosong University, Senior Digital Expert and Ph.D. Student, Korea

* Corresponding author: Jun Cui (Email: jcui228@student.solbridge.ac.kr)

1. Introduction

As China's rural revitalization strategy continues to advance, the role of agricultural finance has become increasingly critical. The agricultural finance supply chain is a key component of rural revitalization, as it provides the necessary capital support for the sustainable development of agriculture, thereby ensuring the achievement of rural revitalization goals. This chain involves not only the financing stages of agricultural production but also the subsequent stages such as procurement, storage, transportation, and processing, all of which require funding. However, the agricultural supply chain is characterized by its length, numerous stages, multiple stakeholders, and low levels of centralization, meaning its funding needs are often fragmented, small, and short-term, which brings a series of risks. Among these, the issue of commercial credit reallocation is particularly prominent and requires significant attention.

The secondary allocation of commercial credit stems from financial repression in China, where commercial banks dominate the financial system. Due to unresolved "ownership discrimination" issues caused by financial repression and credit discrimination, a substantial amount of bank credit flows to state-owned enterprises or large firms (Wang et al., 2015; Chen et al., 2020). Consequently, large enterprises, such as publicly listed companies, provide financing support to small and medium-sized enterprises (SMEs) along the supply chain through commercial credit channels, thereby easing SMEs' financing constraints (Ayaggari et al., 2010; Cunat, 2007; Giannetti et al., 2011). However, the nature of commercial credit reallocation involves enterprises acting as credit intermediaries and participating in shadow banking activities (Harford et al., 2014; Duchin et al., 2017; Zhang & Liu, 2018), a process that carries significant leverage and opacity, introducing a degree of risk.

The impact of secondary commercial credit allocation on corporate finances is dual-faceted. On one hand, it increases financial and management costs, as well as bad debt losses, elongating the supply chain of credit funds and amplifying financial risks

*across the supply chain (Buchak et al., 2018; Allen et al., 2019). On the other hand, the repayment risk of commercial credit demanders may reduce their own solvency, transmitting through accounting accounts to commercial credit suppliers, thus escalating risk contagion among upstream and downstream supply chain enterprises, leading to risk accumulation (Li & Han, 2019). Such risk contagion, when an issue arises with any entity in the supply chain, further exacerbates systemic financial risks (Zhang, 2019).*

*Moreover, the information asymmetry between banks and SMEs lowers SMEs' access to loans, creating opportunities for large firms to engage in secondary commercial credit allocation (Coulibaly et al., 2013). Efforts to curtail this reallocation are essential to limit the role of large firms in commercial credit. By leveraging big data credit assessment, banking fintech can effectively identify the credit status of SMEs, channeling more credit to the prioritized areas of agriculture, rural communities, and SMEs, thus reducing the motivation for large firms to engage in secondary commercial credit reallocation, while increasing agricultural finance input (Kong et al., 2021). In this context, banking fintech has begun to play an increasingly important role in promoting innovation and exploration in agricultural supply chain finance.*

*Banking fintech refers to the application of emerging technologies in banking operations, including artificial intelligence, blockchain, cloud computing, big data, and the Internet. In recent years, more commercial banks have begun adopting fintech in their business processes to enhance service efficiency and innovation capabilities (Shahrokhi, 2008; Tan & Wang, 2023). The application of banking fintech, previously focused primarily on urban centers, is now extending into agriculture, driving the online, digital, and scenario-based development of agricultural supply chain finance (Berg et al., 2020). Through the integration of advanced technologies, banks can achieve digital management and monitoring of the agricultural supply chain, improving risk identification and response capabilities (Cheng & Qu, 2020). For instance, big data technology can provide comprehensive monitoring of agricultural production and sales stages, promptly identifying and addressing potential risk factors. Additionally, banking fintech enables quick access to SMEs' credit information (Goldstein et al., 2019), reducing the costs associated with information search and risk control (Vasiljeva et al., 2016), thereby curbing credit risk (Cheng & Qu, 2020) and lowering financing costs (Wonglimpiyarat, 2019), ultimately enhancing the financing efficiency of agricultural supply chain finance and providing robust support for the stable development of the agricultural supply chain. Moreover, fintech offers a new positioning for agricultural supply chain finance, as it typically revolves around the purchase of receivables, replacing traditional financial service models through*

*disruptive innovation (Lee & Yong, 2018) and linking core enterprises with upstream and downstream entities to establish trust mechanisms and ensure transaction security (Fuster et al., 2019). This approach constructs a financing model that flexibly provides financial products and services, integrating fund flows with supply chain management, optimizing fund accessibility and cost (Berger & Udell, 1995), and mitigating risks arising from information asymmetry (Livshits et al., 2016).*

*In practice, banks use core enterprises in the supply chain as starting points (Kshetri, 2018), providing funding to financially constrained upstream suppliers to prevent supply chain collapse or offering bank credit to downstream distributors to solidify long-term supply chain partnerships with core enterprises (Berger et al., 2005). This stability and reliability in partnerships reduce the likelihood of secondary commercial credit allocation. By digitizing management, utilizing extensive agricultural credit information, and constructing comprehensive agricultural credit assessment models, banking fintech effectively inhibits secondary commercial credit allocation in agricultural supply chains, thereby bolstering the stable development of these chains. This application not only facilitates business expansion and risk management for banks but also fosters the development and innovation of the agricultural supply chain.*

*The preceding discussion highlights the key role of banking fintech in addressing supply chain financial risks. Currently, game theory is receiving attention as an effective tool for studying the relationship between bank credit and supply chain financial risk. Scholars are constructing game models to analyze the participatory behavior of economic agents, seeking effective ways to address SMEs' financing challenges. In this field, Li & Xin (2017) explored the impact of supply chain finance on SME financing by contrasting traditional credit and supply chain credit. Mao & Zhu (2016) constructed an evolutionary game model to analyze the application of supply chain finance in traditional logistics, offering relevant recommendations and decision-making insights to guide future development. Considering that the bounded rationality assumption of evolutionary game models more closely aligns with actual agents' behavior, Sheng & Chen (2019) established a dynamic evolutionary model involving the government, logistics enterprises, and financial institutions to analyze the evolutionary dynamics of supply chain finance. Additionally, Wang et al. (2022) developed a dynamic evolutionary model featuring financial institutions, core enterprises, and SMEs to study financing pathways under supply chain financial credit, exploring the evolution of the supply chain finance credit system toward new financing models in which core enterprises provide credit guarantees for SMEs. Broadening financing channels for SMEs has been a focal point for scholars, and Lu (2014)*

*discovered, based on an analysis of bank, core enterprise, and SME dynamic game dynamics, that a stable supply chain relationship enables banks to better understand SMEs, thereby assisting in obtaining credit. Our study is closely related to the works of Sheng & Chen (2019) and Wang et al. (2022), as both analyze ways to expand SMEs' funding availability within an evolutionary game framework. However, unlike previous studies, we focus on the financial risks arising from secondary commercial credit allocation. This study investigates how banking fintech enhances SMEs' access to bank credit by reducing secondary commercial credit allocation and maintaining supply chain financial security. This unique perspective provides new theoretical foundations for understanding how banking fintech mitigates agricultural supply chain financial risks.*

*Although topics on fintech and supply chain financial risks have been widely discussed (Yan et al., 2020; Sung & Shirley, 2020), there is a lack of in-depth research on the specific impact of banking fintech on supply chain financial risk, especially within the evolutionary game framework of cooperation and competition among various parties in the agricultural financial supply chain. Further exploration is needed regarding the micro-level mechanisms through which banking fintech affects agricultural supply chain financial risk. Therefore, this paper seeks to investigate the mechanisms through which banking fintech operates by analyzing the gaming behavior among economic agents, specifically considering the uniqueness of agricultural financial supply chains. It aims to explore how banking fintech, by influencing the behaviors and interests of core enterprises and SMEs within the supply chain, suppresses agricultural supply chain financial risks, thereby addressing gaps in existing literature.*

*Based on bounded rationality, this study constructs a three-player evolutionary game model involving banks, core enterprises, and SMEs in the supply chain to examine the impact of banking fintech on agricultural supply chain financial risk. By optimizing the agricultural financial supply chain and incentivizing cooperation among banks, core enterprises, and SMEs, this research aims to contribute to achieving rural revitalization objectives and sustainable agricultural economic development while providing references for relevant policy decisions.*

*Our findings indicate that banking fintech applications play a pivotal role in reducing both financing costs and financial risks for core enterprises and small and medium-sized enterprises (SMEs) within the agricultural supply chain. By providing advanced digital solutions, such as big data credit assessment, blockchain-backed transaction security, and AI-driven risk evaluation, banking fintech enhances both the reliability and efficiency of financial transactions, minimizing the costs associated with*

*traditional lending processes.*

2. Literature review

Modern agricultural supply chains operate within a multi-layered ecosystem of stakeholders, from raw material procurement to final product dispatch across multiple geographical locations, forming a non-linear and non-sequential chain. With the evolution of the times and the adjustment of agricultural structure, China's agriculture is facing many challenges and opportunities. On the one hand, this is closely related to China's unique national conditions. Specifically, China's population accounts for about 22% of the world's population and its cultivated land area accounts for about 7% of the world's (Smil, 1995; Brown and Halweil, 1998; Liu and Lu, 2001). Agricultural enterprises are faced with low levels of automation and poor management skills. , information inconsistency and fragmentation, product adulteration, and food safety issues are serious challenges (Luthra et al., 2018; Ritchie et al., 2018; Singh et al., 2019). On the other hand, with the continuous advancement of urbanization and the further improvement of people's living standards, China has made remarkable achievements in increasing grain production (Zhao et al., 2008).

As the economy continues to develop, consumer demand for agricultural products continues to increase, which puts tremendous pressure on agriculture and natural resources, exacerbating the instability of the agricultural supply chain, and posing a major threat to the sustainable development of agriculture (Zhao et al. al., 2008). In addition, it should be noted that issues such as payment delays, high transaction costs, and extended delivery times also pose threats to the security of agricultural supply chain finance (Balaji and Arshinder, 2016). In this context, agricultural supply chain finance (SCF) has become particularly critical and is directly related to national food security and the sustainable development of the rural economy. As a method of coordinating the behavior of upstream and downstream organizations in the supply chain, optimizing capital flows, and creating value, agricultural supply chain finance has become an effective means to maintain the financial security of agricultural supply chains and promote rural revitalization (Hofmann, 2005).

Stemmler and Seuring (2003) first proposed the term "supply chain finance". After development and evolution, Hofmann (2005) defined it as a method of "co-creating value through planning, guidance and control of financial resources on the supply chain". Subsequently, Gomm (2010) further developed this concept and believed that supply chain finance can optimize the financial status of enterprises in the supply chain. He emphasized the role of information technology in supply chain finance and put

*forward the perspective of integrating with information technology. . However, over time, Bryant and Camerinelli (2013) pointed out that the definition of supply chain finance has gradually become blurred, driven by financial market development and technological changes. This may be attributed to a combination of factors, including rapid technological change, changing market demands, developments in financial markets, and the nature of supply chain finance as an interdisciplinary field. Overall, the definition of supply chain finance has become richer and more diverse over time, reflecting its adaptability in evolving business and technological environments. At the same time, the global financial crisis has also shaped the evolution of supply chain finance to a certain extent. (Hofmann and Belin, 2011). The 2008 crisis prompted central banks across the world to take action and inject large amounts of liquidity, but this did not benefit all enterprises equally, resulting in small and medium-sized enterprises, especially agricultural enterprises, still facing serious financing problems (Moretto and Caniato, 2021). It is in this context that supply chain finance emerges as a new financing model, whose main goal is to improve access to funds and management of financial flows throughout the supply chain, focusing on supporting those weaker players (Gelsomino et al. , 2016), especially agricultural-related enterprises that have been most significantly affected. However, despite its positive intentions, supply chain finance also faces some major challenges in the process of sustainable development (Jia et al., 2020). Specifically, for those suppliers who are in a weak position in the supply chain, due to lack of funds and the inability to afford the high interest rates brought by bank credit, financial barriers are considered to be the main constraints that restrict such suppliers from participating in sustainable development. factors (Birkin et al., 2009; Glover et al., 2014). At this time, suppliers with financing advantages make secondary allocations of commercial credit by utilizing the excess credit funds obtained by financial privileges (Harford et al., 2014). According to the resource reallocation theory, customers are usually required to pay insurance premiums and default premiums in order to Obtain superior financial returns (Cunat, 2007). This behavior will lead to adverse effects on downstream customers, including lower supplier participation and insurance premiums and default premiums brought about by the secondary allocation of commercial credit, thus seriously affecting the efficiency and sustainability of the entire supply chain. Therefore, it is urgent to formulate and implement plans to promote the sustainable development of supply chain finance (Carter and Easton, 2011; Gopalakrishnan et al., 2012).*

*The core function of the financial system is to convert savings into investment and promote the effective allocation of resources (Levine, 1997). However, in China, this function is restricted by long-standing institutional problems such as financial*

*repression and "ownership discrimination" (Chen et al., 2020). In addition, small and medium-sized enterprises generally face the lack of collateral and detailed financial information. challenges, making them face difficulties when seeking financing from traditional commercial banks (Beck and Demirguc-kunt, 2006), especially agricultural-related enterprises, due to special challenges such as seasonal risks, long-term return cycles, market uncertainty and difficulty in asset pledge , are more susceptible to financing difficulties, which will lead to interruptions in the production chain, endangering the production and circulation of agricultural products, and thus threatening the security of the agricultural supply chain. State-owned enterprises and large listed enterprises with financing advantages use their financial privileges to engage in shadow banking business (Duchin et al., 2017), providing bank credit funds and funds obtained from the stock market to downstream small, medium and micro enterprises in the form of commercial credit. The secondary allocation of commercial credit plays a role as a credit intermediary in the financial system to a certain extent, and plays a role in easing the capital supply pressure of small and medium-sized enterprises (Du et al., 2017). However, although the secondary allocation of commercial credit alleviates the capital supply pressure of small and medium-sized enterprises, it also causes some potential problems, such as reducing market information transparency, exacerbating information asymmetry among economic entities, and affecting the efficiency of capital allocation ( Allen et al., 2019). The secondary allocation of commercial credit is actually an informal financial activity. The lack of supervision makes the operation of the financial system more complicated. These problems not only affect the long-term sustainable development of enterprises, but also pose potential threats to the overall security of supply chain finance. Therefore, it is crucial to seek a balance to ensure the robustness of the financial system and the security of supply chain finance (Jia et al., 2020).*

*In recent years, the rapid development of financial technology (FinTech) has become the focus of the international financial field. The Financial Stability Board (FSB) defined FinTech in 2016 as technology-driven financial innovation, with the main goal of improving financial services through technological innovation. Globally, the rapid rise of financial technology has attracted widespread attention from academic circles, especially in the financial market. Emerging financial technologies have played a positive role in improving convenience, accelerating transaction speed, and improving security (Begenau et al. al., 2018; Fuster et al., 2019; Zhu, 2019). However, with the rapid development of financial technology, the banking industry has also been profoundly affected by it (He et al., 2022; Begley and Srinivasan, 2022). Bank financial technology refers to the application of emerging technologies in the banking industry,*

*such as artificial intelligence, blockchain, cloud computing and big data, to improve service quality and efficiency. In China, commercial banks have adopted financial technology to improve loan technology, speed up information transmission, reduce credit risks, and promote corporate innovation. With the development of bank fintech, a lot of literature on bank fintech has emerged in the academic community (Cenni et al., 2015; Liberti and Petersen, 2018; Sheng, 2021; Chen and Yang, 2020; Tan et al., 2023 ).*

*Bank fintech plays a key role in solving problems such as misallocation of traditional financial funds and financial suppression. Unlike traditional banks that provide old-fashioned, costly and cumbersome financial services (Brandl and Hornuf, 2017), bank financial technology widely uses applications such as Internet information technology, big data, blockchain technology and artificial intelligence when providing credit to small and medium-sized enterprises. (Jagtiani and Lemieux, 2017). This effectively helps banks improve the availability and accuracy of information, increase information channels and sources, and reduce information friction between banks and SMEs (Athreya et al., 2012; Sedunov, 2017; Sanchez, 2018). For agricultural companies, this means more convenient and efficient financial services, which will help improve the financing availability and credit ratings of agricultural enterprises. The application of bank financial technology can also help improve banks' ability to process risk information and reduce information processing costs (DeYoung et al., 2011; Livshits et al., 2016). This is particularly critical for the agricultural field, because agricultural production and operations are often affected by external factors such as weather and natural disasters, and more flexible and rapid financial support is needed to deal with risks. Among agriculture-related companies, small, medium and micro enterprises are often the main business entities. Since there are significant differences in risk and credit qualifications between these enterprises and large enterprises, there is an obvious market segmentation in their financing sources, which provides an opportunity for enterprises with financing convenience to use them to obtain low-cost excess credit funds for secondary allocation of commercial credit. conditions (Petersen and Rajan, 1997). The application of bank financial technology will have an indirect impact on the secondary allocation of commercial credit of listed and large enterprises by improving credit funding support for small, medium and micro enterprises that are subject to greater financing constraints. Specifically, the inhibitory effect of bank financial technology on the secondary allocation of commercial credit will be manifested through the demand side and supply side.*

*On the one hand, the development of bank financial technology has produced significant technological spillover effects on the supply side, helping to alleviate the information asymmetry between upstream and downstream small and medium-sized*

*enterprises and banks in the agricultural supply chain (Sheng et al., 2021). Taking China as an example, the construction of Alipay's rural credit system provides a new credit assessment method for agricultural-related enterprises. By analyzing multi-dimensional data such as transaction data and agricultural production information, it can improve credit trust in agricultural-related enterprises. Through financial technology, banks can more accurately assess and monitor the credit of agricultural-related enterprises, reduce credit risks (Blalock and Gertler, 2008; Newman et al., 2015; Cheng et al., 2020), and then more actively provide services to agricultural-related enterprises. Enterprises provide financing, improve the efficiency of credit approval, and increase credit supply (Sheng et al., 2021), thereby easing the credit distortion caused by financial repression, reducing financing costs, and weakening the incentives of listed companies to provide commercial secondary credit. Driven by bank fintech, this more comprehensive and accurate credit assessment of agricultural-related enterprises also reduces the phenomenon of "ownership discrimination" and enables a more equitable distribution of financial resources. Compared with the traditional financial system, fintech platforms pay more attention to data The objectivity and comprehensiveness of the system reduce credit discrimination caused by corporate ownership forms, increase the difficulty for listed companies to obtain excess credit funds, further weaken the ability of secondary allocation of commercial credit, thereby reducing supply chain financial risks.*

*On the other hand, according to the substitution theory, agricultural-related companies, as demanders of commercial credit, are usually in a relatively weak position. Large enterprises such as listed companies will ask for insurance premiums and default premiums from customers when providing commercial credit (Cunat, 2006), which leads to an increase in their cost of obtaining commercial credit. Agricultural-related companies face a more difficult situation due to demand-side restrictions. . Relying on advanced technologies such as big data, artificial intelligence and blockchain, bank fintech has been successfully applied in complex and multi-level supply chains. This not only improves the agricultural production process and makes the supply chain more agile and elastic, but also enables Meet seasonal and cyclical financing needs and improve economic benefits, thereby supporting the comprehensive development of digital agriculture (Mukherjee et al., 2021). Specifically, this technology-driven financial environment has significantly improved the level of credit assessment (Chen and Yang, 2020), improved information transparency between banks and agricultural supply chain companies, and companies can have a clearer understanding of the relationship between banks and business providers. The difference in financing costs between credit financings allows them to weigh the pros and cons more wisely when*

*making financing decisions. Compared with commercial credit with higher commercial credit costs, they prefer to obtain bank loans and reduce financing costs (Ng et al., 1999 ; Wilner, 2000). Therefore, driven by the motivation to reduce financing costs, small, medium and micro enterprises are more willing to seek bank financing and reduce their reliance on intermediary financing companies, which directly affects the structure of their financing needs and inhibits the secondary allocation of commercial credit from the demand side.*

*To sum up, the development of financial technology has greatly alleviated the demand for high financing cost commercial credit by agricultural-related enterprises and brought opportunities for the sustainable development of the agricultural supply chain.*

*The rapid development of financial technology (fintech) has had a profound impact on the agricultural sector, particularly in addressing the high costs associated with traditional commercial credit. Historically, agricultural enterprises have faced challenges in securing affordable and accessible financing due to the high-risk nature of the industry, which often results in high financing costs. These challenges are exacerbated by the agricultural supply chain's vulnerability to price volatility, seasonal demand fluctuations, and natural disasters. However, the emergence of financial technology has significantly altered this dynamic by providing more efficient and lower-cost alternatives for financing. This shift has the potential to enhance the sustainability and resilience of the agricultural supply chain, which is essential for ensuring food security and promoting economic growth in rural areas.*

*Financial technology, with its ability to leverage big data, machine learning, and blockchain technology, has enabled the creation of innovative financial products and services tailored to the unique needs of the agricultural sector. These advancements have led to improved risk assessment models, more accurate credit scoring, and faster loan disbursement processes. By reducing transaction costs and improving access to capital, fintech has made it easier for agricultural enterprises to obtain the necessary funding without relying on traditional commercial credit channels, which are often expensive and restrictive.*

*This transformation not only benefits individual agricultural enterprises but also has broader implications for the entire agricultural supply chain. One key area where fintech has proven advantageous is in reducing the secondary allocation of commercial credit. In traditional financing models, intermediaries such as banks and other financial institutions often play a significant role in the allocation of credit, leading to increased costs and inefficiencies. The integration of financial technology disrupts this model by directly connecting agricultural businesses with lenders and investors,*

*thereby eliminating the need for multiple intermediaries. This reduction in the secondary allocation of credit helps to streamline financing processes, lower overall costs, and improve the efficiency of capital distribution within the supply chain.*

*Moreover, by facilitating more efficient financial transactions and providing greater transparency, fintech can contribute to reducing financial risks within the agricultural supply chain. Agricultural supply chains are inherently risky due to factors such as crop failure, market fluctuations, and supply disruptions. Financial technology tools, such as predictive analytics and blockchain-based smart contracts, offer greater visibility into supply chain activities and improve risk management practices. These tools enable more accurate forecasting, ensure better coordination among supply chain participants, and facilitate timely payments, all of which help to reduce the financial uncertainties that have traditionally plagued the agricultural sector.*

*Based on these observations, we hypothesize that, when other conditions remain constant, the integration of bank financial technology helps to mitigate the secondary allocation of commercial credit within the agricultural supply chain, thereby reducing financial risks. By providing more direct and efficient access to financing, fintech enhances the stability and sustainability of agricultural enterprises, ultimately contributing to a more resilient agricultural supply chain. This hypothesis is grounded in the theory of financial inclusion and the risk-reducing benefits of technological innovation in financial services, which emphasize how fintech can democratize access to capital, reduce reliance on traditional credit channels, and lower the overall financial risk within vulnerable sectors like agriculture.*

3. *Theory fondations.*

*Building upon the theoretical foundations outlined in the literature review, this study utilizes the following theories to substantiate and support the research framework.*

*Firstly, **Financial Inclusion Theory** provides a critical lens through which the role of financial technology (fintech) in enhancing access to financial services within the agricultural sector is understood. By emphasizing the reduction of barriers to financing, this theory supports the argument that fintech platforms can democratize financial access for agricultural enterprises, especially smallholder farms that traditionally face high financing costs and limited credit options. The integration of fintech solutions is expected to empower agricultural businesses, enabling them to access affordable capital and improve their operational sustainability.*

*Secondly, **Risk Reduction and Technological Innovation Theory** plays a pivotal role in understanding how fintech innovations can mitigate the financial and*

*operational risks associated with agricultural supply chains. The theory highlights that through advanced technologies such as big data, machine learning, and blockchain, fintech can significantly reduce the uncertainties that characterize agricultural markets. By enhancing risk assessment and decision-making processes, fintech helps agricultural enterprises avoid costly commercial credit, thus lowering financial risk and fostering stability within the supply chain.*

*Furthermore,* **Efficiency and Intermediary Reduction Theory** *supports the hypothesis that fintech solutions can optimize financial processes by eliminating traditional intermediaries. In the context of agricultural financing, where access to credit has often been impeded by intermediaries such as banks and financial institutions, fintech platforms streamline the credit allocation process. This disintermediation reduces transaction costs and enhances the speed of financial flows within the agricultural supply chain, improving overall efficiency and enabling more effective resource allocation.*

*In addition,* **Supply Chain Finance Theory** *is employed to examine how fintech can enhance the financial management of agricultural supply chains. By facilitating smoother financial transactions, improving liquidity, and reducing financial friction, fintech ensures that agricultural businesses can better manage cash flows, even during periods of financial instability. The application of fintech in supply chain finance further reduces payment delays, enhances transparency, and strengthens the resilience of the agricultural supply chain.*

*Lastly,* **Innovation Diffusion Theory** *is utilized to explain the adoption process of fintech solutions within the agricultural sector. According to this theory, the widespread adoption of fintech is driven by its perceived advantages, such as reduced costs, increased accessibility, and compatibility with existing business practices. The theory posits that once early adopters within the agricultural community experience the benefits of fintech, its adoption will spread more broadly, further transforming the financial landscape of agriculture.*

*Together, these theories form a robust conceptual framework that supports the hypothesis that the adoption of fintech can reduce the secondary allocation of commercial credit in the agricultural supply chain, thereby mitigating financial risks and promoting the sector's sustainable development.*

*Financial Inclusion Theory*

*Financial inclusion theory posits that access to financial services is a fundamental driver of economic empowerment, particularly in underserved sectors such as agriculture. The theory emphasizes that when individuals and businesses gain access*

*to affordable and reliable financial products, their economic prospects improve significantly. In the context of agricultural enterprises, which have historically struggled with high financing costs and limited access to traditional credit, the development of financial technology (fintech) serves as a critical enabler of financial inclusion. By utilizing digital platforms, agricultural businesses can bypass traditional intermediaries, such as commercial banks, that often impose high transaction fees and stringent lending criteria. This democratization of financial services allows agricultural enterprises, particularly smallholder farmers, to access the capital they need to expand their operations, invest in innovation, and manage risks effectively. The theory suggests that fintech not only facilitates easier access to capital but also fosters a more inclusive financial environment that can enhance the long-term sustainability of the agricultural sector.*

*Risk Reduction and Technological Innovation Theory*

*Risk reduction and technological innovation theory explores how the application of advanced technologies can mitigate various forms of business and financial risks, especially in high-risk industries like agriculture. This theory underscores the importance of leveraging innovative tools, such as big data analytics, machine learning algorithms, and blockchain technology, to improve risk management and decision-making processes. In agricultural supply chains, these technologies enable more accurate credit scoring, enhance the predictability of financial outcomes, and reduce the uncertainty associated with fluctuating agricultural prices and climate-related disruptions. By integrating fintech solutions, agricultural enterprises can better assess and manage financial risks, minimizing the need for costly traditional commercial credit, which often exacerbates these risks due to high interest rates and limited flexibility. Fintech platforms also allow for greater transparency and real-time monitoring of transactions and supply chain activities, which further reduces the risk of fraud, defaults, and other financial irregularities. Therefore, the theory argues that fintech innovation plays a crucial role in mitigating financial risks, particularly in sectors where volatility and uncertainty are prevalent.*

*Efficiency and Intermediary Reduction Theory*

*The efficiency and intermediary reduction theory focuses on the idea that technological advancements can streamline business processes by reducing the number of intermediaries involved in transactions. In traditional financing models, agricultural enterprises often rely on banks and other financial institutions to secure funding, which entails numerous intermediaries, each adding costs and delays to the process. With the advent of fintech, this traditional model is disrupted, as digital platforms directly connect agricultural enterprises with lenders, investors, and other financial service*

*providers. This disintermediation reduces the cost of credit and enhances the efficiency of capital allocation within the agricultural supply chain. According to this theory, fintech reduces inefficiencies inherent in traditional financial systems by eliminating unnecessary intermediaries and enabling faster, more transparent transactions. This efficiency not only benefits individual enterprises but also contributes to the overall stability and sustainability of the agricultural supply chain, as resources are allocated more effectively and at a lower cost. Consequently, the theory highlights the transformative power of fintech in fostering a more efficient and competitive financial ecosystem for the agricultural sector.*

*Supply Chain Finance Theory*

*Supply chain finance theory examines how financial solutions can optimize the flow of capital throughout the supply chain, from suppliers to producers and end consumers. The theory emphasizes the role of financial technology in facilitating smoother transactions and reducing the financial friction that typically arises in supply chains, especially in industries like agriculture, where payment cycles are long, and liquidity is often constrained. In the context of agricultural supply chains, fintech offers tools such as blockchain-based smart contracts and digital invoicing systems, which enhance transparency, speed up payment cycles, and reduce payment delays. These technological innovations improve cash flow management for agricultural businesses, enabling them to maintain operations even during periods of financial strain. Furthermore, fintech solutions can help mitigate the impact of market volatility by providing real-time access to market data, enabling businesses to make more informed decisions about pricing, purchasing, and sales strategies. Thus, the theory suggests that supply chain finance solutions, powered by fintech, can significantly reduce financial risk and improve the overall resilience of agricultural supply chains.*

*Innovation Diffusion Theory*

*Innovation diffusion theory, as proposed by Rogers (2003), explains how new technologies are adopted and spread within a social system over time. According to this theory, the rate of adoption of an innovation depends on various factors, including its relative advantage, compatibility with existing practices, complexity, trialability, and observability. In the context of fintech, the theory suggests that agricultural enterprises, especially small and medium-sized farms, are likely to adopt fintech solutions if these technologies offer clear advantages, such as lower costs, greater accessibility, and easier integration into existing business operations. The diffusion of fintech innovations within the agricultural sector can be accelerated by demonstrating their effectiveness in improving financial access, reducing risks, and enhancing supply chain efficiency. As more agricultural businesses observe the benefits of fintech adoption, they are more*

*likely to follow suit, creating a positive feedback loop that fosters widespread adoption and drives the digital transformation of the agricultural sector. Thus, innovation diffusion theory highlights the critical role of early adopters and the gradual spread of fintech solutions in transforming the financial landscape of agriculture.*

*These theories collectively provide a comprehensive framework for understanding how financial technology can reduce the financial risks faced by agricultural enterprises, streamline credit allocation processes, and contribute to the long-term sustainability and growth of agricultural supply chains.*

4. Methodology

Model Construction

*In recent years, as financial activities in the agricultural sector have increased, the issue of secondary allocation of commercial credit within the agricultural supply chain has become increasingly severe. This phenomenon not only raises borrowing costs for agricultural SMEs but also lengthens the agricultural capital supply chain, thereby amplifying the overall financial risks within the agricultural supply chain (Buchak et al., 2018; Allen et al., 2019). In light of the financing challenges and high financial risks faced by agricultural SMEs, it has become essential to mitigate these risks and provide more accessible financing channels for SMEs to foster the growth of the rural economy. Reducing financial risks in the agricultural supply chain will inevitably suppress the secondary allocation of commercial credit. This process involves key stakeholders: banks, core enterprises (denoted as Enterprise A) that provide commercial credit funds, and SMEs (denoted as Enterprise B) that require financing. Due to the divergent interests among these three parties, the strategies they employ to address the secondary allocation of commercial credit will dynamically shift based on each other's actions.*

*This paper, set against the backdrop of reducing the agricultural supply chain's financial risks and curbing the secondary allocation of commercial credit, utilizes evolutionary game theory to analyze the strategic behaviors of these various stakeholders as banks improve their financial technology capabilities. Additionally, this study examines the stable equilibrium points of the three-party strategies by constructing a dynamic three-party evolutionary game model.*

*Consider an agricultural supply chain comprising Enterprise A and Enterprise B. In the context of suppressing the secondary allocation of commercial credit, the interaction of strategies adopted by the three stakeholders can be illustrated as follows:*

*When addressing the reduction of financial risks in the agricultural supply chain*

*and curbing the secondary allocation of commercial credit, banks are inclined to enhance their financial technology. The improvement in financial technology enables banks to better identify and manage risks (Cheng & Qu, 2020), reduce information search costs, and lower the risk control costs involved in providing funds to enterprises (Vasiljeva et al., 2016). This, in turn, enhances the efficiency and capacity of banks to provide financing and reduces credit risks in the financing process (Cheng & Qu, 2020). Therefore, when the investment required to enhance financial technology is outweighed by the benefits it brings to the bank, banks will opt to improve their financial technology to offer financing channels to SMEs.*

*For Enterprise A, when the bank enhances its financial technology, Enterprise A is unlikely to offer commercial credit funds. As a rational decision-maker, Enterprise A must weigh its financial advantages in providing commercial credit to SMEs and securing excess returns (Harford et al., 2014; Cunat, 2007) against the risks associated with such funding, such as long capital return cycles or even the possibility of no returns due to market uncertainties. Thus, in the context of suppressing the secondary allocation of commercial credit, if the costs and risks for Enterprise A to provide commercial credit exceed the potential returns, it will tend not to engage in providing such credit.*

*For Enterprise B, when the bank improves its financial technology, it is more likely to opt for bank financing. With the enhancement of financial technology, banks are able to offer SMEs more access to information and technology such as big data and internet platforms during the financing process (Jagtiani & Lemieux, 2017), thereby reducing information asymmetry between Enterprise B and the bank (Sedunov, 2017; Sanchez, 2018). This reduction in information friction leads to lower credit risks and financing costs for Enterprise B when choosing to finance through banks. Additionally, the improvement in bank financial technology facilitates the financing process for agricultural SMEs, making it easier for them to gain bank credit approval (Sheng et al., 2021). Therefore, when banks enhance their financial technology, if the costs and risks of financing through banks are lower than the costs associated with commercial credit from Enterprise A, Enterprise B is more likely to choose bank financing.*

1. *Problem Analysis*

    *Table: Parameters of the Three-Party Game Model*

| **Subjects** | **Participating Interests** | **Description** |
| --- | --- | --- |
| ***Bank*** | *Rgf* | *Initial revenue of the bank* |
| | *Cg* | *Investment cost for the bank to improve financial technology* |

|  |  | Cgf | Cost of bank loans before enhancing financial technology |
|---|---|---|---|
|  |  | m | Interest rate on bank loans |
| Enterprise A |  | e | Financing interest rate of Enterprise A |
|  |  | Cm | Mediation fees charged by Enterprise A |
|  |  | Caf | Borrowing costs for Enterprise A |
| Enterprise B |  | I | Financing amount needed by Enterprise B |
|  |  | u | Probability of Enterprise B repaying the bank |
|  |  | v | Probability of Enterprise B repaying Enterprise A |
|  |  | w | Probability of Enterprise B obtaining financing from the bank |
|  |  | Cbf | Credit guarantee cost for Enterprise B when financing from the bank |

2. Model Hypothesis

Assumption 1: In a scenario where other constraints are not considered, the bank, Enterprise A, and Enterprise B are treated as a unified system. Within this system, all three parties are assumed to be decision-makers with bounded rationality. Each decision is made independently based on complete information available at the time. Furthermore, the strategic choices of these parties are influenced by the outcomes of previous interactions, meaning that over time, their strategies gradually evolve and ultimately stabilize at an optimal equilibrium.

Assumption 2: If the bank opts to enhance its financial technology level, it will result in a reduction of the credit costs for Enterprise B when seeking financing from the bank. Conversely, if the bank does not improve its financial technology, Enterprise B will incur a certain credit guarantee cost when obtaining financing from the bank. Alternatively, if Enterprise B chooses to seek financing from Enterprise A, it will incur a specific intermediary cost. Regardless of whether the financing is sourced from the bank or from Enterprise A, Enterprise B will face labor costs associated with securing the funds. Furthermore, after securing financing, the likelihood of Enterprise B repaying either the bank or Enterprise A is denoted by the probabilities $u$ and $v$, respectively.

*Assumption 3: When Enterprise B borrows from either the bank or Enterprise A, it will be subject to an interest rate of m and e, respectively. Additionally, if Enterprise B chooses to obtain financing from the bank, the probability that the bank will approve the loan is denoted by w, which depends on the creditworthiness of Enterprise B.*

*Assumption 4: In the three-party evolutionary game involving the bank, Enterprise A, and Enterprise B, the strategy set for the bank is denoted as S1={improve,not improve}, where x represents the probability that the bank will choose to improve its financial technology, and 1−x is the probability that it will not. The strategy set for Enterprise A is S2={provide commercial credit funds,not provide commercial credit funds}, where y is the probability that Enterprise A will provide commercial credit funds, and 1−y is the probability that it will not. The strategy set for Enterprise B is S3={choose bank financing,choose enterprise A financing}, where z denotes the probability that Enterprise B will choose to finance through the bank, and 1−z is the probability that it will opt for financing from Enterprise A.*

*Three-party payment matrix*

Table **2**.  Payoff Matrix

| Items | Banks improve their financial technology level (x) | | Banks do not improve their financial technology level (1-x) | |
|---|---|---|---|---|
| | Enterprise A provides commercial credit funds (y) | Enterprise A does not provide commercial credit funds (1-y) | Enterprise A provides commercial credit funds (y) | Enterprise A does not provide commercial credit funds (1-y) |
| **Enterprise B chooses bank financing (z)** | Rgf+(u*m*I+(1-u)*0)-Cg | Rgf+(u*m*I+(1-u)*0)-Cg | Rgf+(u*m*I+(1-u)*0)-Cgf | Rgf+(u*m*I+(1-u)*0)-Cgf |
| | 0 | 0 | 0 | 0 |
| | w*(I-m*I)+(1-w)*0 | w*(I-m*I)+(1-w)*0 | w*(I-m*I)+(1-w)*0-Cbf | w*(I-m*I)+(1-w)*0-Cbf |
| **Enterprise B chooses Enterprise A for** | Rgf-Cg | Rgf-Cg | Rgf | Rgf |
| | -Caf+(v*e*I+(1-v)*0)+Cm | 0 | -Caf+(v*e*I+(1-v)*0)+Cm | 0 |

| | | | | | |
|---|---|---|---|---|---|
| commercial financing (1-z) | I-Cm-e*I | 0 | I-Cm-e*I | 0 | |

## 5. Results

Model analysis

### 3.1 Replication Dynamic Analysis

Assuming that E11, E12 and Ex represent the expected benefits of the banking sector improving its financial technology level, the expected benefits of not improving its financial technology level and the average expected benefits, we hypothesize as following:

E11=y*z*(Rgf+(u*m*I+(1-u)*0)-Cg)+(1-y)*z*(Rgf+(u*m*I+(1-u)*0)-Cg)+y*(1-z)*(Rgf-Cg)+(1-y)*(1-z)*(Rgf-Cg)

E12=y*z*(Rgf+(u*m*I+(1-u)*0)-Cgf)+(1-y)*z*(Rgf+(u*m*I+(1-u)*0)-Cgf)+y*(1-z)*Rgf+(1-y)*(1-z)*Rgf

Ex=x*E11+(1-x)*E12

According to Malthus's dynamic equation, the following banking sector replication dynamic equation is derived:

Fx=x*(E11-Ex)=-x*(x*(y*(Cg-Rgf)*(z-1)+y*z*(Rgf-Cg+I*m*u)-(Cg-Rgf)*(y-1)*(z-1)-z*(y-1)*(Rgf-Cg+I*m*u))+(x-1)*(Rgf*y*(z-1)-y*z*(Rgf-Cgf+I*m*u)-Rgf*(y-1)*(z-1)+z*(y-1)*(Rgf-Cgf+I*m*u))-y*(Cg-Rgf)*(z-1)-y*z*(Rgf-Cg+I*m*u)+(Cg-Rgf)*(y-1)*(z-1)+z*(y-1)*(Rgf-Cg+I*m*u))

Similarly, the average expected return and replication dynamic equation of Enterprise A and Enterprise B can be obtained:

Enterprise A: E21, E22, Ey represent the expected return of Enterprise A providing commercial credit funds, the expected return of not providing commercial credit funds, and the average expected return respectively:

E21=x*z*0+x*(1-z)*(-Caf+(v*e*I+(1-v)*0)+Cm)+(1-x)*z*0+(1-x)*(1-z)*(-Caf+(v*e*I+(1-v)*0)+Cm)

E22=x*z*0+x*(1-z)*0+(1-x)*z*0+(1-x)*(1-z)*0

Ey=y*E21+(1-y)*E22

Fy=y*(E21-Ey)=y*(y*(x*(z-1)*(Cm-Caf+I*e*v)-(x-1)*(z-1)*(Cm-Caf+I*e*v))-x*(z-1)*(Cm-Caf+I*e*v)+(x-1)*(z-1)*(Cm-Caf+I*e*v))

Enterprise B: E31, E32, and Ez represent the expected return of Enterprise B if it chooses to obtain financing through a bank, the expected return of Enterprise A's commercial channel financing, and the average expected return:

E31=x*y*(w*(I-m*I)+(1-w)*0)+x*(1-y)*(w*(I-m*I)+(1-w)*0)+(1-x)*y*(w*(I-m*I)+(1-w)*0-Cbf)+(1-x)*(1-y)*(w*(I-m*I)+(1-w)*0-Cbf)

E32=x*y*(I-Cm-e*I)+x*(1-y)*0+(1-x)*y*(I-Cm-e*I)+(1-x)*(1-y)*0

Ez=z*E31+(1-z)*E32

Fz=z*(E31-E3z)=z*((y*(x-1)*(Cm-I+I*e)-x*y*(Cm-I+I*e))*(z-1)+z*((Cbf-w*(I-I*m))*(x-1)*(y-1)-y*(Cbf-w*(I-I*m))*(x-1)-w*x*y*(I-I*m)+w*x*(I-I*m)*(y-1))-(Cbf-w*(I-I*m))*(x-1)*(y-1)+y*(Cbf-w*(I-I*m))*(x-1)+w*x*y*(I-I*m)-w*x*(I-I*m)*(y-1))

3.2 Balanced Strategy Analysis

According to the replication dynamic equation of the bank, enterprise A and enterprise B, a three-dimensional dynamic system can be constructed:

let (Fx, Fy, Fz) = (0, 0, 0), The following equilibrium points can be obtained: E1[0,0,0], E2[1,0,0], E3[0,1,0], E4[1,1,0], E5[0,0,1], E6[1,0,1], E7[(Cbf-Cm+I-I*e-I*w+I*m*w)/Cbf, 1, Cg/Cgf], E8[(Cbf-I*w+I*m*w)/Cbf, 0, Cg/Cgf]。

The equilibrium point may be located at a fixed point in the game strategy space, or at the boundary and interior of the strategy space. The equilibrium point located inside the strategy space may be a central point or a saddle point, not an ESS of the evolutionary game. Therefore, this paper mainly studies the first six pure strategy equilibrium points. On the basis of the above six equilibrium points, the asymptotic stability of the above six equilibrium points is further determined by the local stability of the Jacobi matrix. The Jacobi matrix of the three-dimensional dynamic system is:

$$J = \begin{bmatrix} F_x(x) & F_y(x) & F_z(x) \\ F_y(y) & F_y(y) & F_y(y) \\ F_z(z) & F_z(z) & F_z(z) \end{bmatrix}$$

Table Asymptotic stability of local equilibrium points

| Equilibrium point | Character value | Result |
| --- | --- | --- |
| [0,0,0] | $\lambda 1 = -Cg$<br>$\lambda 2 = Cm-Caf+I*e*v$<br>$\lambda 3 = I*w-Cbf-I*m*w$ | when Cm-Caf+I*e*v<0, I*w-Cbf-I*m*w<0 It is a stable point when, otherwise it is a saddle point or an unstable point |
| [1,0,0] | $\lambda 1 = Cg$ | when Cm-Caf+I*e*v>0, |

|  | $\lambda 2 = Cm-Caf+I*e*v$<br>$\lambda 3 = I*w-I*m*w$ | $I*w-Cbf-I*m*w>0$ is an unstable point, otherwise it is a saddle point |
| --- | --- | --- |
| [0,1,0] | $\lambda 1 = -Cg$<br>$\lambda 2 = Caf-Cm-I*e*v$<br>$\lambda 3 = Cm-Cbf-I+I*e+I*w-I*m*w$ | when $Caf-Cm-I*e*v<0$, $Cm-Cbf-I+I*e+I*w-I*m*w<0$ It is a stable point when, otherwise it is a saddle point or an unstable point |
| [1,1,0] | $\lambda 1 = Cg$<br>$\lambda 2 = Caf-Cm-I*e*v$<br>$\lambda 3 = Cm-I+I*e+I*w-I*m*w$ | **When** $Caf-Cm-I*e*v>0$, $Cm-I+I*e+I*w-I*m*w>0$ It is an unstable point when, otherwise it is a saddle point |
| [0,0,1] | $\lambda 1 = Cgf-Cg$<br>$\lambda 2 =0$<br>$\lambda 3 = Cbf-I*w+I*m*w$ | **When** $Cgf-Cg<0$, $Cbf-I*w+I*m*w<0$ It is a stable point when, otherwise it is an unstable point or saddle point |
| [1,0,1] | $\lambda 1 = Cg-Cgf$<br>$\lambda 2 =0$<br>$\lambda 3 = I*m*w-I*w$ | **When** $Cg-Cgf<0$, $I*m*w-I*w<0$ It is a stable point when, otherwise it is an unstable point or a saddle point |

*Scenario 1: when $Cm-Caf+I*e*v<0$, $I*w-Cbf-I*m*w<0$ 时, E1[0,0,0] The stable equilibrium is characterized by the bank choosing not to improve its financial technology level, enterprise A refraining from providing commercial credit funds, and enterprise B opting for financing from enterprise A. In this scenario, the bank decides against upgrading its financial technology because the required investment for such improvements may outweigh the potential benefits it could bring. Similarly, enterprise A lacks the incentive to offer commercial credit financing to enterprise B, as the financing costs and the risk of non-repayment by enterprise B exceed the potential returns from providing credit. This outcome is primarily driven by the issues of information asymmetry and credit risk between the bank and the financing parties.*

*Scenario 2: when $Caf-Cm-I*e*v<0$, $Cm-Cbf-I+I*e+I*w-I*m*w<0$, E3[0,1,0] Stable Point Scenario: In this scenario, the banking sector chooses not to enhance its financial technology level, enterprise A provides commercial credit funds, and enterprise B opts for commercial financing from enterprise A. The bank has no incentive to invest in improving its financial technology, as the investment costs may outweigh the potential benefits. Consequently, the bank maintains its current strategy. Enterprise A, on the other hand, is motivated to provide commercial credit funds to enterprise B,*

*as doing so allows enterprise A to earn interest income and intermediary fees, which surpass the costs associated with offering the credit. Enterprise B chooses to secure commercial credit financing from enterprise A because, although this option incurs certain intermediary costs, it is more favorable than financing from the bank. Given that the bank has not improved its financial technology, information asymmetry exists between the bank and enterprise B. If enterprise B were to approach the bank for financing, it would face credit guarantee costs and the risk of loan rejection. Therefore, enterprise B finds it more advantageous to seek financing from enterprise A.*

*Scenario 3: when $C_{gf}-C_g<0$, $C_{bf}-I*w+I*m*w<0$, E5[0,0,1] Stable Point Scenario: In this case, the banking sector chooses not to enhance its financial technology capabilities, enterprise A does not provide commercial credit funds, and enterprise B opts to secure financing from the bank. The bank refrains from improving its financial technology because the required investment may outweigh the potential benefits it could bring. Similarly, enterprise A has no incentive to provide commercial credit funds to enterprise B, as the costs associated with offering the credit and the risks of enterprise B defaulting on repayment exceed the potential income from providing such funds. On the other hand, while the intermediary costs for enterprise B to secure financing from enterprise A may exceed the credit guarantee costs and credit risks associated with bank financing, enterprise B ultimately chooses to approach the bank for funding.*

*Scenario 4: when $C_g-C_{gf}<0$, $I*m*w-I*w<0$, E6[1,0,1] Stable Point Scenario: In this case, the banking sector enhances its financial technology capabilities, enterprise A does not provide commercial credit funds, and enterprise B opts to secure financing from the bank. The bank chooses to improve its financial technology because the ability to more efficiently obtain enterprise information and reduce credit risk between the bank and enterprises results in substantial income that outweighs the necessary investment costs. Consequently, the bank invests in enhancing its financial technology. Enterprise B chooses to finance through the bank, as the improved financial technology reduces information asymmetry and credit risk, resulting in a lower cost for enterprise B compared to seeking commercial credit from enterprise A. Enterprise A, on the other hand, refrains from offering commercial credit, as the associated costs and risks outweigh the potential benefits. Furthermore, with the bank's enhanced financial technology, enterprise B is more inclined to seek financing from the bank, leaving enterprise A with limited opportunities to generate profits.*

*4. Numerical simulation*

*The previous analysis of the equilibrium strategies and replicator dynamics*

equations of the three parties clearly identifies the potential stable equilibrium points. Through a systematic examination of the relevant sectors, we have gained insights into the key factors influencing their strategies. Based on empirical research data and theoretical analysis, we have determined the following parameters.: $I=10$, $R_{gf}=0$, $C_g=1$, $u=0.85$, $m=0.2$, $C_{af}=1$, $C_m=1.5$, $v=0.8$, $e=0.25$, $w=0.8$, $C_{gf}=1$, $C_{bf}=1$。 The simulation step size is set to... 20

4.1 The impact of fintech investment costs on the strategies of the three parties.（$C_g$）

To examine the impact of fintech investment costs on the strategies of the three parties, we define the proportion of fintech investment costs in the ESS as... $C_g=$（1.0，1.5，2.0）。The simulation results are presented in the figure below. Figure 2 illustrates the impact of rising fintech investment costs on the probability that the banking sector will enhance its fintech capabilities. As shown in Figure 2(a), with the increase in fintech investment costs, the benefits derived by the bank from fintech fail to cover the associated investment costs, leading to a decreased likelihood of the bank improving its fintech level. In scenarios where the fintech investment cost is high, the bank opts not to enhance its fintech capabilities. As a result, the information asymmetry and risks between the bank and small and medium-sized enterprises (SMEs) grow, leading to frequent reallocation of commercial credit funds in the agricultural supply chain. This results in an increased probability of core enterprise A providing commercial credit funds to SMEs, as shown in Figure 2(b). Furthermore, the decline in the probability of banks enhancing fintech capabilities, due to rising fintech costs, also leads to a reduction in the likelihood of SME B choosing bank financing, as depicted in Figure 2(c). In some extreme cases, even though core enterprise A provides commercial credit financing to SMEs, the high intermediary fees prompt SMEs to prefer bank financing instead, as shown in Figure 2(d).

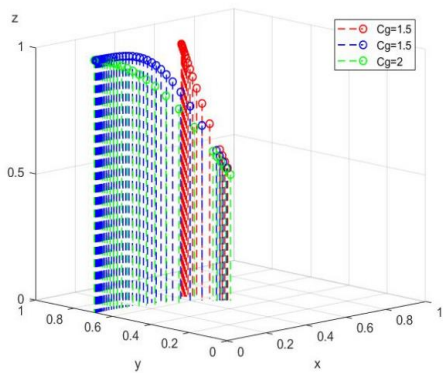

Figure 2（a）

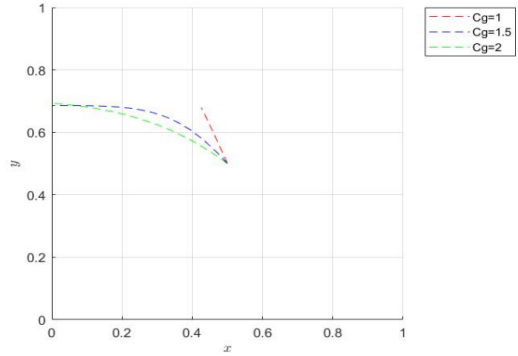

Figure 2（b）

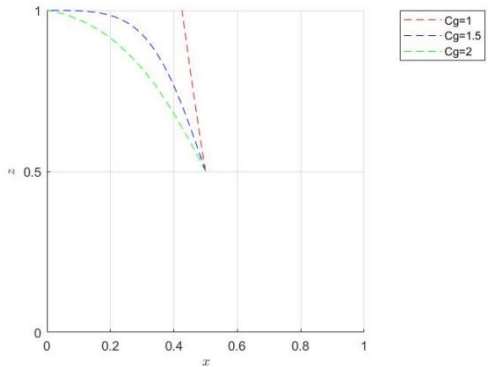

Figure 2（c）

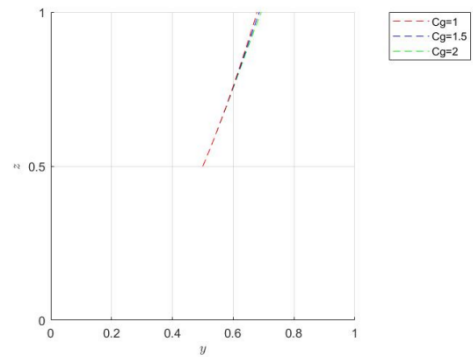

Figure 2（d）

4.2 The Impact of Loan Interest Rate Levels on the Strategies of the Three Parties（m，e）

To examine the impact of loan interest rates on the strategies of the three parties, the bank loan interest rate in the ESS is set as follows:m=（0.20，0.25，0.30），

The commercial credit loan interest rates for enterprise A are set at e = (0.25, 0.30, 0.35). The specific simulation results are shown in Figures 3 and 4. As depicted in Figures 3(a) and (c), an increase in the bank loan interest rate significantly reduces the probability of SME B choosing bank financing. The rise in bank loan interest rates leads to a substantial increase in the financing costs for SMEs, which can exceed the benefits they receive. As a result, SME B shifts its preference to obtaining commercial credit financing from enterprise A.

In contrast, the impact of changes in the bank loan interest rate on enterprise A's strategic decisions is relatively minimal, as shown in Figure 3(b). Moreover, as the bank loan interest rate increases, the probability of enterprise A providing commercial credit financing also rises. When the bank loan interest rate is sufficiently high, SME B alters its behavior and opts for commercial credit financing from enterprise A instead. This shift results in a negative correlation between the probability of SME B choosing bank financing and the probability of enterprise A offering commercial credit financing, as

*shown in Figure 3(d).*

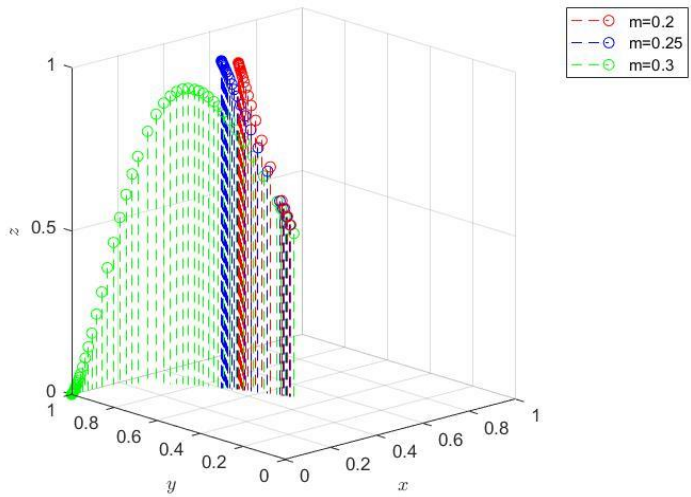

*Figure 3（a）*

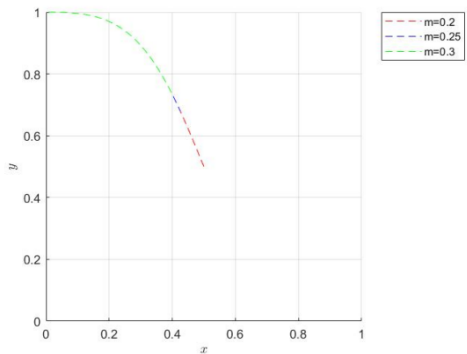 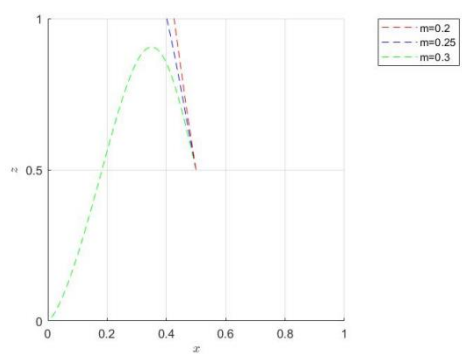

*Figure 3（b）* *Figure 3（c）*

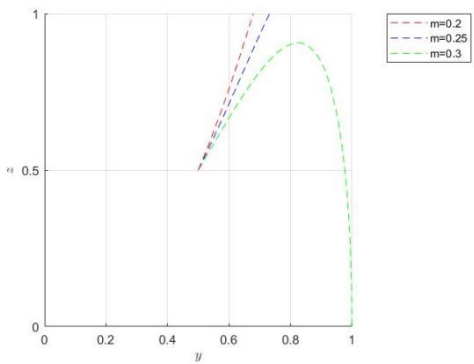

*Figure 3（d）*

*The impact of changes in the commercial credit loan interest rate on the strategies of the three parties is relatively small, as shown in Figures 4(a) and 4(b). However, an increase in the commercial credit loan interest rate slightly raises the probability of enterprise B choosing bank financing, as illustrated in Figure 4(c). This is because the*

increase in the commercial credit loan interest rate raises the cost for enterprise B to obtain financing from the core enterprise. Figure 4(d) shows that the rise in the commercial credit loan interest rate slightly increases enterprise A's willingness to provide commercial credit funds, as the potential returns for enterprise A increase. However, the impact on enterprise B's strategy is minimal.

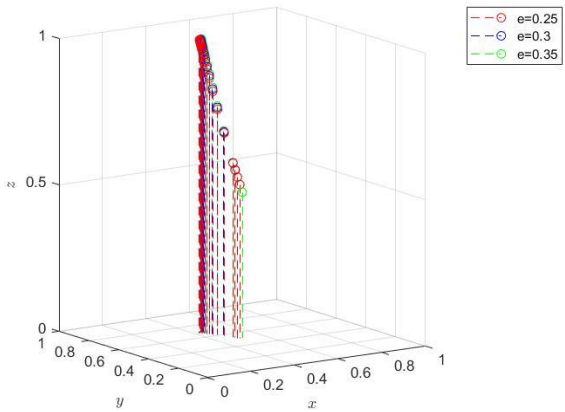

Figure 4（a）

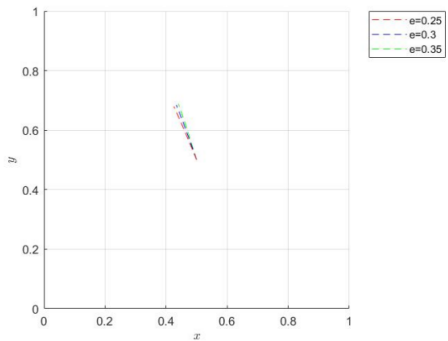 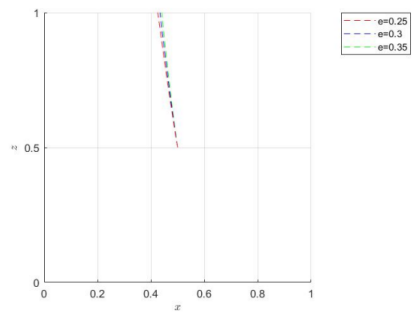

Figure 4（b）                                    Figure 4（c）

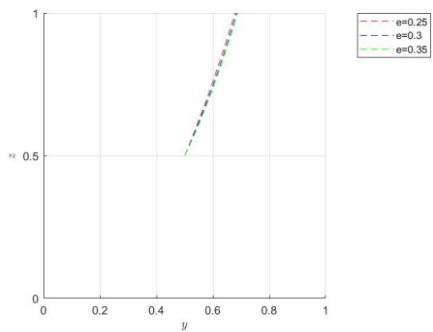

Figure 4（d）

4.3 The Impact of Commercial Credit Financing Intermediary Costs on the Strategies of the Three Parties（Cm）

To examine the impact of intermediary costs in commercial credit financing on the strategies of the three parties, we set the intermediary cost in ESS as $Cm = (1.5, 2.0, 2.5)$. The specific simulation results are shown in Figure 5. Figure 5 illustrates the impact of intermediary costs in commercial credit financing on the strategies of the three parties, assuming a constant level of financial technology.

The increase in commercial credit financing costs raises the cost for enterprise B to use the commercial credit channel, which slightly increases the probability of enterprise B choosing bank financing, as shown in Figures 5(a) and 5(c). At the same time, the rise in intermediary costs has a modest impact on the bank's decision to improve its financial technology level, as shown in Figure 5(b). As intermediary costs increase, more small and medium-sized enterprises choose to finance through banks. To mitigate credit risks and enhance profitability, banks are more likely to increase their financial technology investment.

Furthermore, the increase in intermediary costs enhances the financing returns for enterprise A, which slightly increases the probability of enterprise A providing commercial credit funds, as illustrated in Figure 5(d).

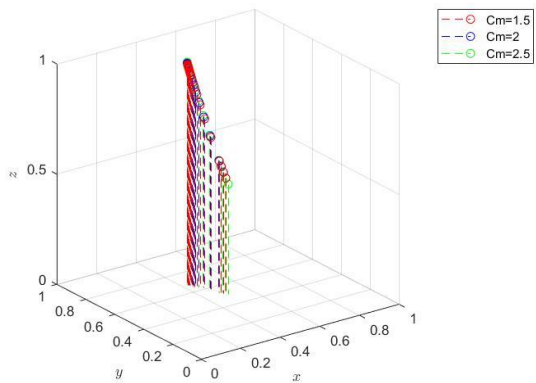

Figure 5（a）

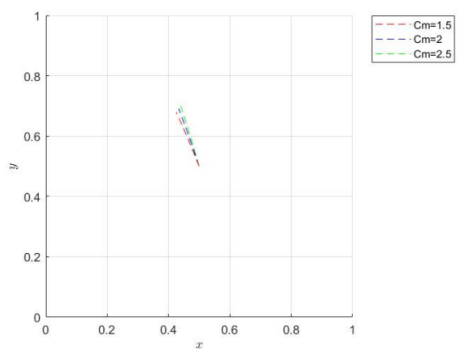

Figure 5（b）

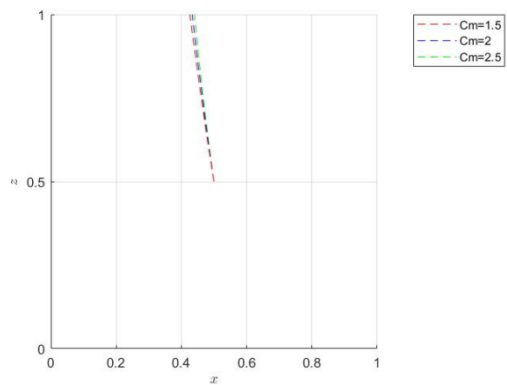

Figure 5（c）

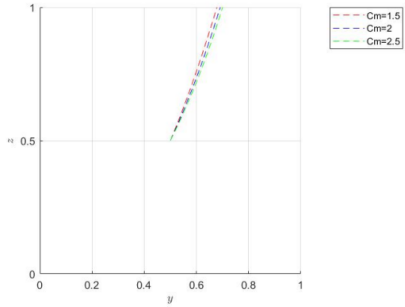

*Figure 5（d）*

The Impact of Enterprise Financing Amount (I) on the Strategies of the Three Parties

To investigate the impact of the financing amount on the strategies of the three parties, we set the financing amount in ESS as I = (10, 12, 14). The specific simulation results are presented in Figure 6.

As shown in Figure 6(a), the curve indicates that the financing amount has a relatively minor effect on the strategies of the three parties. However, the curve also demonstrates that as the financing amount increases, the probability of the bank improving its financial technology level and the probability of enterprise A providing commercial credit funds both rise. This is because larger financing amounts lead to higher potential returns for both the bank and enterprise A, thus motivating more active participation in their respective strategies.

Moreover, the increase in financing amount also enhances the likelihood that enterprise B will choose bank financing, as shown in Figures 6(c) and 6(d). When the financing requirement of enterprise B is larger, the credit risks and costs associated with bank financing are relatively lower compared to using the commercial credit channel. As a result, enterprise B is more inclined to select bank financing.

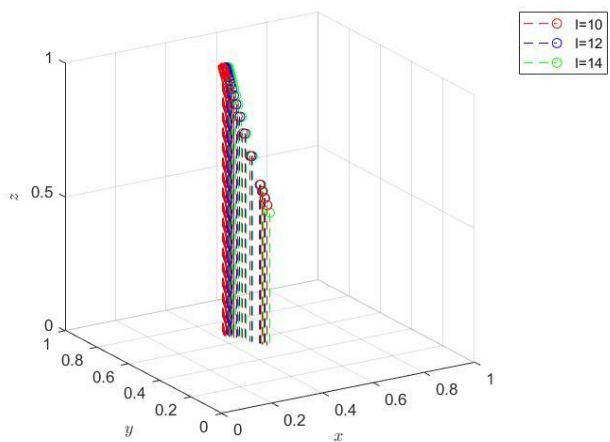

*Figure 6（a）*

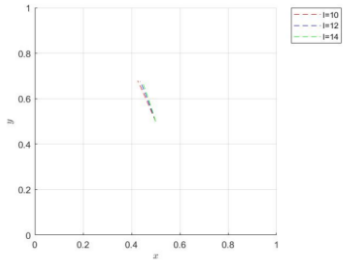
Figure 6（b）

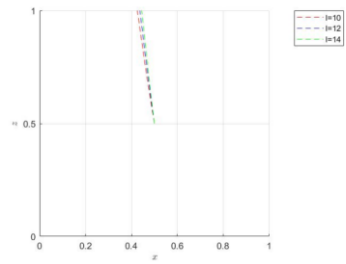
Figure 6（c）

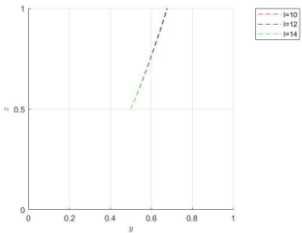
Figure 6（d）

6. Disussion and Conclusion

*To promote rural revitalization comprehensively, it is essential to strengthen financial services for rural development, particularly in supporting agriculture and poverty alleviation areas. Agricultural loans are a vital source of funding for rural revitalization. In recent years, financial services for rural revitalization and other agricultural sectors have gradually strengthened, with the banking sector continually expanding the coverage, accessibility, and balance of financial services for "agriculture, rural areas, and farmers" (referred to as "San Nong"). By 2022, the outstanding balance of agricultural loans, both domestic and foreign currency, reached 49.25 trillion yuan, a year-on-year increase of 14%, surpassing the previous year's growth rate by 3.1 percentage points. Under national policies that guide services for the "San Nong," financial institutions have made steady progress in agricultural sector investments. As the rural revitalization strategy advances, the key role of agricultural finance has become increasingly prominent. The agricultural financial supply chain is a crucial link that cannot be ignored and is a fundamental guarantee for achieving rural revitalization goals. This supply chain covers not only the financing support for agricultural production but also the subsequent funding needs in stages such as procurement, storage, and processing. However, due to the complexity of the agricultural supply chain, involving multiple links and entities, and the small scale and low degree of centralization of individual enterprises, the funding demands are fragmented, small-scale, and short-term, leading to a series of risks within the agricultural supply chain. A particularly significant issue is the problem of secondary*

*allocation of commercial credit.*

*Secondary allocation of commercial credit refers to enterprises acting as credit intermediaries and engaging in shadow banking activities (Harford et al., 2014; Duchin et al., 2017), channeling funds from bank loans to support financing needs of upstream and downstream small and medium-sized enterprises (SMEs) in the supply chain through commercial credit channels (Ayaggari et al., 2010; Cunat, 2007; Giannetti et al., 2011). This has provided substantial support in alleviating the financing difficulties of SMEs, but it also increases financial costs, management costs, and even bad debt losses, resulting in longer credit supply chains and exacerbating the overall financial risks in the supply chain (Buchak et al., 2018; Allen et al., 2019). While increasing the investment in agricultural finance, financial institutions are also accelerating the exploration and innovation of agricultural supply chain finance. In this context, bank financial technology (Bank FinTech) has become increasingly important.*

*Bank financial technology refers to the application of emerging technologies in banking operations, including artificial intelligence, blockchain, cloud computing, big data, and the internet. In recent years, more commercial banks have started to adopt financial technologies in their business processes to improve service efficiency and innovation capacity (Shahrokhi, 2008; Tan and Wang, 2023). The application of bank financial technology is not only widespread in urban centers but is also expanding into the agricultural sector (Berg et al., 2020), promoting the digitalization, online expansion, and scenario-based development of agricultural supply chain finance. By introducing advanced technologies, banks can achieve digital management and monitoring of agricultural supply chains, improving their ability to identify and respond to risks (Cheng and Qu, 2020). For example, big data technology allows for comprehensive monitoring of agricultural production and sales processes, enabling the timely identification and resolution of potential risks. Furthermore, through financial technology, banks can quickly assess the credit status of SMEs (Goldstein et al., 2019), reducing information search and risk management costs (Vasiljeva et al., 2016), thereby controlling credit risk (Cheng and Qu, 2020) and reducing financing costs (Wonglimpiyarat, 2019). This enhances the financing efficiency of agricultural supply chain finance and provides strong support for the stable development of agricultural supply chains. Additionally, the application of financial technology offers a new positioning for agricultural supply chain finance. Traditionally, supply chain finance is based on accounts receivable factoring, but financial technology, through disruptive innovations (Lee and Yong, 2018), breaks past technological limitations. Banks can link core enterprises with upstream and downstream companies, establish trust mechanisms,*

*ensure transaction security and reliability (Fuster et al., 2019), and create a flexible financing model that integrates cash flow and supply chain management, optimizing fund availability and costs (Berger and Udell, 1995). This reduces the risks arising from information asymmetry (Livshits et al., 2016). Specifically, banks can offer financing to upstream suppliers facing difficulties or provide bank credit to downstream distributors, thereby stabilizing the long-term cooperation between core enterprises and downstream distributors (Berger et al., 2005). The stability and reliability of these cooperation relationships reduce the likelihood of secondary allocation of commercial credit. Through digital management and the exploration of multi-channel, wide-coverage agricultural credit information, bank financial technology builds a more comprehensive agricultural credit assessment model, accurately evaluates credit conditions (Khandani et al., 2010), and establishes trust mechanisms (Livshits et al., 2016). This effectively suppresses secondary allocation of commercial credit within the agricultural supply chain, providing strong support for its stable development.*

*Although the topic of financial technology and supply chain financial risks has received widespread attention (Yan et al., 2020; Sung and Shirley, 2020), in-depth research is still lacking on the specific impact of bank financial technology on supply chain financial risks, especially in the context of agricultural financial supply chains. The micro-mechanisms of how bank financial technology influences agricultural supply chain financial risks remain underexplored. Therefore, this study aims to explore how bank financial technology, by affecting the behavior and interests of core enterprises and upstream and downstream SMEs in the agricultural financial supply chain, clarifies the mechanism of action of bank financial technology through an analysis of the game behaviors among economic entities, thereby shedding light on the channels through which it reduces agricultural supply chain financial risks. Based on bounded rationality, this paper constructs a tripartite evolutionary game model involving banks, core enterprises, and upstream and downstream SMEs in the supply chain, analyzing the impact of bank financial technology on agricultural supply chain financial risks. This research will provide insights into optimizing agricultural financial supply chains, incentivizing cooperation among banks, core enterprises, and SMEs, and promoting the realization of rural revitalization and sustainable agricultural economic development.*

*This study explores the impact of banking fintech on agricultural supply chain financial risks, focusing on the mechanisms of how core enterprises and supply chain businesses interact in the context of fintech applications. However, the research is not without limitations. First, the modeling approach adopted in this study assumes rational decision-making by economic agents, which may not always reflect the full*

*range of human behaviors in real-world financial interactions. Additionally, the empirical data used to validate the model are limited, and broader data sets would enhance the generalizability of the findings. Furthermore, the study predominantly focuses on the Chinese agricultural sector and may not fully account for regional variations or differences in agricultural practices in other parts of the world. Future research could expand the scope of the study by incorporating data from different agricultural regions or countries, allowing for a more comprehensive analysis of the effects of fintech on agricultural supply chains globally.*

*Future research could explore several aspects to further understand the dynamics of banking fintech in agricultural supply chains. One avenue for future work could involve investigating the behavioral factors influencing the adoption and implementation of fintech solutions by core enterprises and suppliers in agriculture. Understanding the role of trust, risk perception, and institutional factors could provide a deeper insight into how these technologies affect decision-making processes. Another promising area for future work is the longitudinal analysis of the impact of fintech on agricultural supply chains over time, to assess the long-term effects on financial risks, supply chain stability, and the overall economic sustainability of agriculture. Additionally, future studies could examine how advances in specific fintech technologies, such as blockchain and AI, are being integrated into agricultural finance and how these innovations shape the evolving landscape of agricultural supply chain management.*

*Overall, this study contributes to the growing body of literature on fintech applications in agriculture and supply chain management by offering a detailed analysis of the impact of banking fintech on agricultural supply chain financial risks. The study introduces a game-theoretic framework that explains the interactions among banks, core enterprises, and supply chain businesses, emphasizing the role of trust and collaboration in reducing financial risks. By focusing on agricultural supply chains, the study addresses a critical gap in the literature concerning the specific risks faced by these supply chains and the role of financial technology in mitigating these risks. The findings offer valuable insights for policymakers, financial institutions, and agricultural businesses looking to leverage fintech to enhance financial risk management and support the development of more sustainable and resilient agricultural supply chains. In conclusion, By examining the interactions between banks, core enterprises, and upstream and downstream businesses, it highlights the importance of trust and collaboration in reducing financial risks. The study uses a game-theoretic model to explain the behavior of economic agents in the context of fintech applications, offering a unique perspective on how technology can improve the stability and efficiency of agricultural supply chains. Despite its contributions, the*

*research has limitations, such as the reliance on limited empirical data and the focus on the Chinese agricultural sector, which may not fully capture global variations. Future research could expand on these findings by exploring behavioral factors influencing fintech adoption and examining the long-term impact of fintech on supply chains. Ultimately, this study contributes to the literature on fintech in agriculture and provides practical implications for policymakers, financial institutions, and businesses looking to leverage technology to foster more sustainable and resilient agricultural systems.*